\documentstyle[aps]{revtex}

\begin{document}
\title{Attack the `ping-pong' protocol without eavesdropping}
\author{Qing-yu Cai}
\address{Wuhan Institute of Physics and Mathematics, The Chinese Academy of Sciences,%
\\
Wuhan, 430071, People's Republic of China}
\maketitle

Bostr$\stackrel{..}{o}$m and Felbinger [1] have presented a ping-pong
communication protocol which allows the information transferred in a
deterministic secure manner. The security of this ping-pong protocol is
based on an entangled pair of qubits. And the proof of the case of
eavesdropping attack is correct. The aim of this Comment, however, is to
point out that the information Bob gains from Alice is not reliable, i.e.,
the message can not be transmitted successfully from Alice to Bob if this
`ping-pong' protocol is not modified.

In the `ping-pong' protocol, it utilizes the property that one bit of
information can be encoded in the states $|\psi ^{\pm }>$, which is
completely unavailable to anyone who has only access to one of the qubits.
To gain information from Alice, Bob prepares two qubits in the Bell state $%
|\psi ^{+}>=(1/\sqrt{2})(|0>|1>+|1>|0>)$. Then he stores one qubit and sends
the other one to Alice through the quantum channel. Alice can decide to the
control or the message mode randomly. In message mode, Alice performs a
unitary operation $\sigma _{z}^{A}$ to encode the information `1' or does
nothing to encode the information `0'. Then she sends it back. Bob can get
Alice's information by a Bell measurement. In control mode, Alice performs a
measurement in the basis $B_{z}=\left\{ |0>,|1>\right\} $. Using the public
channel, she sends the result to Bob, who then also switches to control mode
and performs a measurement in the same basis $B_{z}$. Bob compares his own
result with Alice's result. If both results coincide, Bob knows that Eve is
in line and stops the communication. Else, Bob sends next qubit to Alice and
this communication continues. It has been proven that any information Eve
gains would make her face a nonzero detection probability. This is true.
However, a fact that Eve can attack this quantum communication without
eavesdropping is ignored, which would lead to the information Bob gains is
not reliable.

Suppose Eve is in the line. In every control mode, Eve does not touch the
travel back qubit sent by Alice to Bob since Alice announces publicly. In
every message mode, Eve captures the travel back qubit Alice sent to Bob and
performs a measurement in the basis $B_{z}$ and forwards to Bob this qubit.
Alice and Bob have zero probability to find out Eve's attack. Then Bob lets
this communication continue. But, every Bob's measurement result is
meaningless since the two qubits become independent of each other after
Eve's attack measurement. The measurement result of Bob is randomly in
states $|\psi ^{\pm }>$. When the communication is terminated, Bob has
learned nothing but a sequence of nonsense random bits (With zero detection
probability, Eve also can attack every qubit Bob sent to Alice using a
measurement in basis $B_{z}$.).

Essentially, this attack on the `ping-pong' protocol without eavesdropping
is a special case of a Denial-of-Service (DoS) attack. To detect such attack
described above, the `ping-pong' protocol should be modified. Alice can use
a strategy like this: In control mode, with probability $c_{0}$, Alice
measures the travel qubit in basis $B_{z}$ and tells the result to Bob
through public channel. With probability $1-c_{0}$, Alice sends the qubit
back to Bob directly instead of encoding. After Bob receives the qubit (
Maybe Bob should tell Alice about his receipt through public channel in
every message mode. If Alice publishes her operation without Bob's receipt,
Eve would forward the qubit directly. ), she tells Bob about her operation.
If Bob find his measurement result in $|\psi ^{-}>$, there is an Eve in this
line. In this case, Bob has a probability $p=0.5$ to find out Bob's attack.
In fact, this modification can be completely dropped and replaced by any
method of message authentification to protects the protocol against
man-in-the-middle attacks with a reliable public channel [2].

The basis of secure communication is that message can be transmitted
successfully. And the security must be ensured. If the `ping-pong' quantum
communication protocol is not modified, the message transmission cannot be
accomplished under Eve's DoS attack. We can see, such problem does not exist
in BB84 protocol [3], security of which is based on the no-cloning theorem
and that non-orthogonal quantum states cannot be reliably distinguished [4].
Attack without eavesdropping, therefore, aims at the property of the
`ping-pong' protocol.

I thank Yuan-chuan Zou for useful discussions. This work is funded by the
National Science Foundation of China (Grant No. 10004013).

\section{References:}

[1]. K. Bostr$\stackrel{..}{o}$m and T. Felbinger, Phys. Rev. Lett. 89,
187902 (2002).

[2]. B. Schneier, $Applied$ $Cryptography$, 2nd ed. (Wiley, New York, 1996).

[3]. C. H. Bennett and G. Brassard, $Proceedings$ $of$ $the$ $IEEE$ $%
International$ $Conference$ $on$ $Computers$, $Systems$, $and$ $Signal$ $%
Processing$, $Bangalore$ (IEEE, New York, 1984), pp. 175-179.

[4]. M. A. Nielsen and I. L. Chuang, Quantum Computation and Quantum
Information (Cambridge University Press, Cambridge, UK, 2000).

\end{document}